\documentclass{ws-ijmpa}
\usepackage[super,compress]{cite}
\usepackage{graphicx}
\begin{document}

\markboth{G. L.~Klimchitskaya \& V.~M.~Mostepanenko}
{Current status of the problem of thermal Casimir force}

\catchline{}{}{}{}{}

\title{Current status of the problem of thermal Casimir force }

\author{{G.~L.~Klimchitskaya${}^{1,2}$
{{and}}
 V.~M.~Mostepanenko${}^{1,2,3}$}}

\address{${}^1$Central Astronomical Observatory at Pulkovo of the
Russian Academy of Sciences, \\Saint Petersburg,
196140, Russia\\
${}^2$ Peter the Great \\Saint Petersburg
Polytechnic University, Saint Petersburg, 195251, Russia\\
${}^3$Kazan Federal University, Kazan, 420008, Russia\\
g\_klimchitskaya@mail.ru, vmostepa@gmail.com}

\maketitle

\begin{history}
\received{16 December 2021}
\accepted{11 January 2022}
\end{history}

\begin{abstract}
The problem of thermal Casimir force, which consists in disagreement
of theoretical predictions of the fundamental Lifshitz theory with
the measurement data of high precision experiments and some
peculiar properties of the Casimir entropy, is reviewed. We discuss
different approaches to the resolution of this problem proposed in
the literature during the last twenty years. Particular attention is
given to the role of the effects of spatial dispersion. The recently
suggested nonlocal Drude-like permittivities which take proper
account of the dissipation of conduction electrons and bring the
predictions of the Lifshitz theory in agreement with experiment and
requirements of thermodynamics are considered. The prospects of this
approach in the ultimate resolution of the problem of thermal Casimir
force are evaluated.
\keywords{Thermal Casimir force; Lifshitz theory; high precision
experiments; spatial dispersion; Casimir entropy; Nernst heat theorem.}
\end{abstract}

\ccode{PACS Nos.: 12.20.Fv; 12.20.Ds }

\section{Introduction}
\label{secKM:1}
\newcommand{\kb}{{k_{\bot}}}
\newcommand{\skb}{{k_{\bot}^2}}
\newcommand{\vk}{{\mbox{\boldmath$k$}}}
\newcommand{\rv}{{\mbox{\boldmath$r$}}}
\newcommand{\ve}{{\varepsilon}}
\newcommand{\okb}{{(\omega,k_{\bot})}}
\newcommand{\xkb}{{(i\xi_l,k_{\bot})}}
\newcommand{\Mr}{{r_{\rm TM}}}
\newcommand{\Er}{{r_{\rm TE}}}

At zero temperature the Casimir force per unit area of two parallel
ideal metal plates spaced at a distance $a$ (i.e., the Casimir
pressure) is expressed \cite{KM1} as $-\pi^2\hbar c/(240a^4)$. This
force originates from the zero-point fluctuations of quantized
electromagnetic field whose spectra in the presence of planes and in
the free space are different. At nonzero temperature of ideal metal
planes, $T \neq 0$, the Casimir pressure was investigated in several
papers. \cite{KM2,KM3,KM4} At low ($T \ll T_{\rm eff}$) and high
($T \gg T_{\rm eff}$) temperature, where $k_{B}T_{\rm eff} \equiv \hbar c/(2a)$,
it is given by
\begin{equation}
P(a,T)=-\frac{\pi^2\hbar c}{240a^4}
\left[1+\frac{1}{3}\left(\frac{T}{T_{\rm eff}}\right)^4\right],
\qquad
P(a,T)=-\frac{k_{B}T}{4\pi a^3}\zeta(3),
\label{eqKM1}
\end{equation}
\noindent
respectively, where $k_B$ is the Boltzmann constant and $\zeta$(3) is
the Riemann zeta function. These results are determined by the
zero-point and thermal fluctuations of the electromagnetic field.
They are quite reasonable. Specifically, the respective Casimir
entropy per unit area of ideal metal planes goes to zero with
vanishing temperature in line with the Nernst heat theorem. \cite{KM5}.

The Casimir result obtained for ideal metal planes was generalized by
Lifshitz \cite{KM6} for the case of two parallel thick plates
(semispaces) at temperature $T$ described by the frequency-dependent
dielectric permittivity which plays a role of the response function
to electromagnetic fluctuations. At a later time, the Lifshitz theory
was generalized \cite{KM7} for magnetic materials described by the
magnetic permeability. An application region of the Lifshitz theory is
restricted by the fact that it treats the plate material classically
using the continuous permittivity and permeability functions. Furthermore,
the original formulation of this theory takes no account of the
spatial dispersion.

Lifshitz derived his theory on the basis of the fluctuation-dissipation
theorem of quantum statistical physics. This approach is only
applicable to the boundary plates made of usual materials possessing
the dissipation properties. The respective response functions should
be complex, i.e., have some nonzero pure imaginary parts in order the
obtained expressions for the force be nonzero. The reverse is, however,
incorrect. If one substitutes the real dielectric function into the
Lifshitz formula, the obtained Casimir force is not equal to zero.

Another approach to the derivation of the Lifshitz theory is based
on quantum field theory with appropriate boundary conditions. In
doing so, the boundary conditions used by Casimir (the tangential
component of electric field and the normal component of magnetic
induction vanish on the ideal metal planes) are replaced by the
standard continuity conditions of classical electrodynamics. The
resulting formula for the force coincides with the Lifshitz formula
under an assumption of real dielectric permittivity and permeability
functions. \cite{KM8,KM9, KM10} This assumption is needed for
obtaining real energy eigenvalues of the problem. A generalization
of the quantum field theoretical approach to the case of dissipative
materials is reached by considering some auxiliary electrodynamic
problem. \cite{KM11}

Note that the two above approaches are somewhat complementary. The
first of them faces problems for the case of ideal metal planes
(specifically, the Casimir result follows from the Lifshitz formula
by using the special prescription \cite{KM12}). The
fluctuation-dissipation approach is also inapplicable to calculation
of the Casimir energies in some problems of elementary particle
physics (e.g., in the bag model of hadrons \cite{KM13,KM14})
and in cosmological models with nontrivial topology (where the
boundary conditions are replaced with the identification
conditions \cite{KM15,KM16,KM17}). On the other case, as mentioned
above, the second, field-theoretical, approach encounters difficulties
and should be supplemented when applied to the configurations with
boundary bodies made of usual, dissipative, materials. Allowance
must be also made to the fact that in the most of cases the
dielectric functions of condensed matter physics are of more or
less phenomenological character and this may plague an application
of the Lifshitz theory to some specific systems and a comparison
between the measurement data and theoretical predictions.

In this review, we consider the current status of the problem of
thermal Casimir force which was actively discussed in the
literature for the last twenty years. The point is that for usual
metals described by the Drude model the Lifshitz theory predicts
an unexpectedly big thermal correction at short separations
between the plates which decreases the magnitude of attractive
(negative) Casimir force. \cite{KM18} This problem is not
resolved yet because theoretical predictions of the Lifshitz
theory for the room-temperature Casimir force calculated using
the Drude model and similar response functions for both metallic
and dielectric materials turned out to be in conflict with the
measurement data of high precision experiments performed by
different experimental groups. However, the main facts in this
field of research are by now established and some ways to the
resolution of the problem are directed.

The structure of the review is as follows. In Sec.~\ref{secKM:2}, an essence
of the problem of thermal Casimir force between both metallic and
dielectric test bodies is elucidated. Section \ref{secKM:3} presents what is
known from the high precision experiments on measuring the
Casimir interaction. In Sec.~\ref{secKM:4}, different theoretical approaches
to the resolution of this problem proposed during the last years
are briefly considered. Section \ref{secKM:5} is devoted to the question on
whether an alternative response to the evanescent waves could
solve the problem. Finally, Sec.~\ref{secKM:6} contains our conclusions and
outlook.

\section{Problem of the Thermal Correction to the Casimir Force}
\label{secKM:2}

The Lifshitz theory expresses the Casimir force per unit area of material plates
(i.e., the Casimir pressure $P$) as a functional of their dielectric permittivity
$\ve(\omega)$ and magnetic permeability $\mu (\omega)$ which depend on frequency
$\omega$ and may also depend on $T$. For the case of two similar plates the
result is \cite{KM6,KM17}
\begin{eqnarray}
&&
P(a,T)=-\frac{\hbar}{2\pi^2}\int_0^{\infty}\!\!\kb d\kb\int_0^{\infty}\!\!d\omega
\,\coth\frac{\hbar\omega}{2k_BT}
\nonumber \\
&&~~~~
\times{\rm Im}\left\{q\left[r_{\rm TM}^{-2}\okb e^{2aq}-1\right]^{-1}+
q\left[r_{\rm TE}^{-2}\okb e^{2aq}-1\right]^{-1}\right\},
\label{eqKM2}
\end{eqnarray}
\noindent
where
$\kb=(k_1^2+k_2^2)^{1/2}$ is the magnitude of the wave vector projection on
a plane of the plates and $q\equiv q\okb=(\skb-\omega^2/c^2)^{1/2}$.
The transverse magnetic (TM) and transverse electric (TE) reflection coefficients
are defined as
\begin{equation}
\Mr\okb=\frac{\ve(\omega)q-k\okb}{\ve(\omega)q+k\okb},
\qquad
\Er\okb=\frac{\mu(\omega)q-k\okb}{\mu(\omega)q+k\okb}
\label{eqKM3}
\end{equation}
\noindent
and
\begin{equation}
k\okb=\left[\skb-\ve(\omega)\mu(\omega)\frac{\omega^2}{c^2}\right]^{1/2}\! .
\label{eqKM4}
\end{equation}
It is seen that (\ref{eqKM2}) can be represented as the contributions of propagating
waves for which $\omega>c\kb$ and evanescent waves with
$0\leqslant\omega\leqslant c\kb$. The propagating waves satisfy the mass-shell equation in
free space. The evanescent waves are off the mass shell (one can say that they are
characterized by the pure imaginary $k_3$).

Taking into account that the contribution of propagating waves in (\ref{eqKM2}) contains
the rapidly oscillating functions due to the pure imaginary $q$ (but real $k_3$),
in computations it is more convenient to use the mathematically equivalent representation
for $P(a,T)$ in terms of the pure imaginary Matsubara frequencies\cite{KM6,KM17}
\begin{eqnarray}
&&
P(a,T)=-\frac{k_BT}{\pi}\sum_{l=0}^{\infty}{\vphantom{\sum}}^{\prime}
\int_0^{\infty}\!\!q_l\kb d\kb
\left\{\left[r_{\rm TM}^{-2}\xkb e^{2aq_l}-1\right]^{-1}\right.
\nonumber \\
&&~~~~~~~~~~~~~~~~~~~~
+\left.
\left[r_{\rm TE}^{-2}\xkb e^{2aq_l}-1\right]^{-1}\right\}.
\label{eqKM5}
\end{eqnarray}
\noindent
Here, the Matsubara frequencies are $\xi_l=2\pi k_BTl/\hbar$ with
$l=0,\,1,\,2,\,\ldots$, the prime on the summation sign divides the term with $l=0$
by two, and $q_l$ is obtained from $q$ by putting $\omega=i\xi_l$.

The thermal correction to the zero-temperature Casimir pressure is defined as
\begin{equation}
\Delta_TP(a,T)=P(a,T)-P(a,0).
\label{eqKM6}
\end{equation}
\noindent
We consider it separately for metallic and dielectric plates.

\subsection{Thermal correction to the Casimir force between metallic plates}
\label{subsecKM:2p1}

To calculate the Casimir force and the thermal correction to it one needs to know the
response functions of a metal within sufficiently wide frequency region.
In measurements of the Casimir force the nonmagnetic metal Au is in most common use.
The optical data for the
complex index of refraction of Au were measured\cite{KM19} over the range of frequencies
from $\hbar\omega=0.125~$eV to  $\hbar\omega=10^4~$eV making available both
real and imaginary parts of the dielectric permittivity of Au,
${\rm Re}\,\ve_{\rm opt}$ and  ${\rm Im}\,\ve_{\rm opt}$, in the same frequency range.

It should be especially emphasized that the dielectric response is measured in optical
experiments exploiting the propagating waves. It is an assumption that the same response
can be used in the area of evanescent waves as well.

To determine $\ve(i\xi)$ along the imaginary frequency axis, as required in (\ref{eqKM5}),
it is necessary to extrapolate the measured ${\rm Im}\,\ve_{\rm opt}$ to zero frequency
and apply the Kramers-Kronig relation.\cite{KM17}
Under an assumption that the effects of spatial dispersion are sufficiently small
(this is certainly true for the propagating waves), the Maxwell equations result in the first
expansion term of $\ve$ at low frequencies,
$\ve(\omega)=i4\pi\sigma_0/\omega$, where $\sigma_0$ is the conductivity at constant
current.\cite{KM20} This term is well reproduced by the Drude model\cite{KM21}
\begin{equation}
\ve_D(\omega)=1-\frac{\omega_p^2}{\omega[\omega+i\gamma(T)]},
\label{eqKM7}
\end{equation}
\noindent
where $\omega_p$ is the plasma frequency, $\gamma(T)$ is the relaxation parameter and
$\sigma_0=\omega_p^2/(4\pi\gamma)$. The imaginary part of (\ref{eqKM7}) is commonly
used for an extrapolation of the optical data for  ${\rm Im}\,\ve_{\rm opt}(\omega)$ to
$\omega=0$. It was found\cite{KM19,KM22} that for Au $\hbar\omega_p=9.0~$eV and
$\hbar\gamma(300\,\mbox{K})=0.035~$eV.

\begin{figure}[b]
\vspace*{-8.cm}
\centerline{\hspace*{2cm}\includegraphics[width=4.50in]{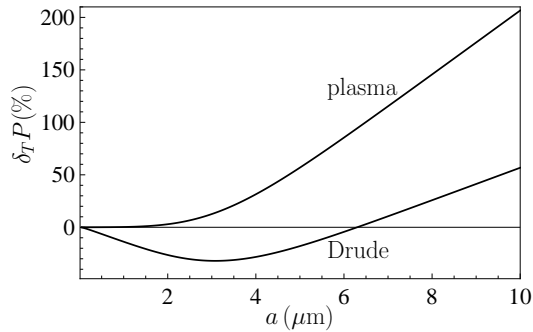}}
\vspace*{-3.7cm}
\caption{The relative thermal correction to the Casimir pressure between Au plates
computed at $T=300~$K using the Drude and plasma models is shown as a function of
separation by the bottom and top lines, respectively.
\protect\label{figKM:1}}
\end{figure}
However, as mentioned in Sec.~\ref{secKM:1}, the Lifshitz theory using the permittivity
(\ref{eqKM7}) predicts rather big thermal correction to the Casimir force which decreases
the force magnitude.\cite{KM18} In Fig.~\ref{figKM:1} the relative thermal correction
$\delta_TP(a,T)=\Delta_TP(a,T)/P(a,0)$ computed by (\ref{eqKM3}) and
(\ref{eqKM5})--(\ref{eqKM7}) at $T=300~$K is shown by the bottom line as a function of
separation between the plates (note that the use of optical data at
$\hbar\omega\geqslant 0.125~$eV makes only a minor impact on the computational results
and only at $a<0.5~\mu$m). As can be seen in Fig.~\ref{figKM:1}, this correction reaches
--6.4\%, --9.4\%, and --13.8\% at $a=0.5$, 0.7, and $1~\mu$m, respectively, and vanishes
at $a=6.3~\mu$m.

The results for Au plates should be compared with respective results for the ideal metal
planes (\ref{eqKM1}). Thus, from the first equality in (\ref{eqKM1}) one obtains that for
ideal metal planes the relative thermal correction takes the positive values, is equal
to only 0.0098\% and 0.157\% at $a=0.5$ and $1~\mu$m, respectively, and increases with
increasing separation. At large separations (high temperatures) for Au plates described
by the Drude model, one obtains one half of the result given by the second
equality in (\ref{eqKM1}) for ideal metal planes.

So strong discrepancy between the thermal corrections predicted theoretically for the
plates made of a good metal and for the ideal metal planes may look incredible.
Taking into account that for two plates spaced at a distance $a$ of the order of
micrometer the characteristic frequency $c/a$  belongs to the region of infrared optics,
it has long been usable to calculate the Casimir force employing the dielectric
permittivity of dissipationless plasma model \cite{KM12,KM23,KM24}
\begin{equation}
\ve_p(\omega)=1-\frac{\omega_p^2}{\omega^2},
\label{eqKM8}
\end{equation}
\noindent
which is obtained from (\ref{eqKM7}) by putting $\gamma(T)=0$.

In Fig.~\ref{figKM:1}, the relative thermal correction
calculated by (\ref{eqKM3}), (\ref{eqKM5}), (\ref{eqKM6}), and (\ref{eqKM8}) is shown by the
top line. Although an application of (\ref{eqKM8}) at low frequencies including the zero
frequency is physically unjustified, the thermal correction shown by the top line
demonstrates the same characteristic properties as in the case of ideal metal planes, i.e.,
it is positive, small at short separations and increases monotonously with increasing
separation. Specifically, at $a=0.5$ and $1~\mu$m the relative thermal correction shown
by the top line is equal to 0.058\% and 0.29\%, respectively. In the high temperature limit,
the thermal Casimir force calculated using the plasma model is given by the second equality
in (\ref{eqKM1}) found in the case of ideal metal planes.

One more fact that deserves attention is regarding an agreement between the Lifshitz theory
and thermodynamics. It was shown that for metals with perfect crystal lattice, which is the
basic idealization of condensed matter physics, the Casimir entropy calculated using the
Drude model (\ref{eqKM7}) violates the third law of thermodynamics (the Nernst heat theorem)
by taking a nonzero (negative) value at $T=0$ depending on the parameters of a
system.\cite{KM25,KM26,KM27,KM28} If, however, the plasma model (\ref{eqKM8}) is used,
the Lifshitz theory results in the zero Casimir entropy at $T=0$, i.e., the Nernst heat
theorem is satisfied.\cite{KM25,KM26,KM27,KM28} It is necessary to stress that in reality
metallic crystal lattices contain some fraction of impurities leading to small nonzero
relaxation at zero temperature $\gamma(0)=\gamma_0$ (for a perfect crystal lattices
$\gamma_0=0$). In this case it was shown that the Casimir entropy jumps to zero at
small $T$ starting from the negative values, i.e., the validity of the Nernst heat
theorem is restored.\cite{KM29,KM30,KM31}

\subsection{Thermal correction to the Casimir force between dielectric plates}
\label{subsecKM:2p2}

As opposite to metals, the dielectric materials possess the zero electric conductivity
at $T=0$. The optical data for dielectric materials, similar to the case of metals, can
be measured over the wide frequency region\cite{KM19} and are basically independent on $T$.
This gives the possibility to obtain their frequency-dependent dielectric permittivity
$\ve_{\rm opt}(\omega)$. In reality, however, dielectric materials possess some nonzero
conductivity $\sigma_0(T)$ at any nonzero temperature which decreases with vanishing $T$
exponentially fast. As a result, the permittivity function of dielectric materials
takes the form \cite{KM19}
\begin{equation}
\ve(\omega)=\ve_{\rm opt}(\omega)+i\frac{4\pi\sigma_0(T)}{\omega}.
\label{eqKM9}
\end{equation}

\begin{figure}[t]
\vspace*{-7.8cm}
\centerline{\hspace*{2cm}\includegraphics[width=4.50in]{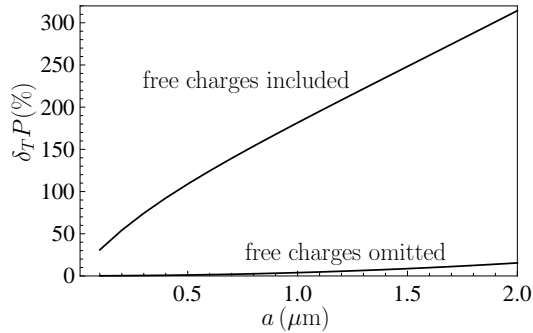}}
\vspace*{-3.8cm}
\caption{The relative thermal correction to the Casimir pressure
between fused silica plates computed at $T=300~$K with included
and omitted free charge carriers is shown as a function of
separation by the top and bottom lines, respectively.
\protect\label{figKM:2}}
\end{figure}

Similar to the metals described by the Drude model, the Casimir force between two dielectric
plates described by
 the permittivity (\ref{eqKM9}) demonstrates an unexpectedly big thermal effect
at short separations. As an example, in Fig.~\ref{figKM:2}  the relative thermal correction
to the Casimir pressure between two fused silica plates computed by (\ref{eqKM3}),
(\ref{eqKM5}) and (\ref{eqKM9}) is shown by the top line as a function of separation at
$T=300~$K. In computations, the values $\ve_{\rm opt}(0)=3.81$ and
$\sigma_0=29.7~\mbox{s}^{-1}$ have been used \cite{KM19} (note that these results are
independent on a specific value of $\sigma_0$, but only on the fact that it is nonzero).
According to Fig.~\ref{figKM:2}, at separations of
1 and $2~\mu$m the relative thermal correction
is equal to 182\% and 314\%, respectively. In the high temperature limit, the thermal
Casimir force between dielectric plates described by (\ref{eqKM9}) is equal to one half of the
result for ideal metals given by the second equality in (\ref{eqKM1}) similar to the case of
metals described by the Drude model (\ref{eqKM7}).

Taking into account that the conductivity of dielectric materials is a very small effect,
it was often omitted in calculations. \cite{KM32,KM33,KM34,KM35,KM36}
In this case the dielectric properties of the plates were characterized by
$\ve_{\rm opt}(\omega)$. The computational results for the relative thermal correction
to the Casimir pressure between two fused silica plates described by $\ve_{\rm opt}(\omega)$
are shown by the bottom line in Fig.~\ref{figKM:2} as a function of separation at
$T=300~$K. For the plates speced
at 1 and $2~\mu$m the relative thermal correction
is equal to  only 3.9\% and 15.4\%, respectively, i.e., much fewer than with included
conductivity. In the limit of high temperature, the Casimir pressure calculated with
omitted conductivity is given by \cite{KM17}
\begin{equation}
P(a,T)=-\frac{k_BT}{8\pi a^3}
{\rm Li}_3\left[\left(\frac{\ve_0-1}{\ve_0+1}\right)^2\right],
\label{eqKM10}
\end{equation}
\noindent
where ${\rm Li}_n(z)$ is a polylogarithm function.

What is more, the use of the dielectric permittivity (\ref{eqKM9}) with included
conductivity of a dielectric material at $T\neq 0$ results in violation  of the Nernst
heat theorem. \cite{KM37,KM38,KM39,KM40,KM41,KM42}
In this case the Casimir entropy at $T=0$ is positive and again depends on the
system parameters. However, if one describes the dielectric materials by
$\ve_{\rm opt}(\omega)$, the Nernst heat theorem is satisfied. \cite{KM37,KM38,KM39,KM40,KM41,KM42}
The Lifshitz theory using the permittivity (\ref{eqKM9}) can be reconciled with the Nernst
heat theorem by assuming that dielectric materials possess some nonzero conductivity down
to zero temperature. \cite{KM43} This is, however, not the case of real dielectrics.

\section{What is Known from High Precision Experiments}
\label{secKM:3}

Several high precision experiments on measuring the Casimir interaction in sphere-plate
geometry have been performed by Ricardo Decca using a micromechanical torsional oscillator
and by Umar Mohideen by means of a modified atomic force microscope (AFM).
All these experiments were done in high vacuum, benefited from direct measurements of the
force-distance relations, and took an advantage of rigorous procedures for comparison between
experiment and theory with no fitting parameters. In this connection, the experiment on
measuring the Casimir force between a spherical lens of centimeter-size curvature radius and
a plate performed by means of the torsion pendulum \cite{KM44} does not fall into the
category of high-precision measurements because a comparison between the measurement data
and theory was made with two fitting parameters.

\subsection{Measurements by means of a micromechanical torsional oscillator}
\label{subsecKM:3p1}

In three subsequent experiments, the gradient of the Casimir force between an Au-coted sphere
of radius $R$ and an Au-coated plate $F_{sp}^{\,\prime}(a,T)$ was measured by means of
a micromechanical torsional oscillator \cite{KM45,KM46,KM47,KM48} and recalculated into
the effective Casimir pressure between two plane parallel plates using the proximity force
approximation (PFA)
\begin{equation}
P(a,T)=-\frac{1}{2\pi R}\,F_{sp}^{\,\prime}(a,T).
\label{eqKM11}
\end{equation}

\begin{figure}[b]
\vspace*{-8cm}
\centerline{
\includegraphics[width=6.50in]{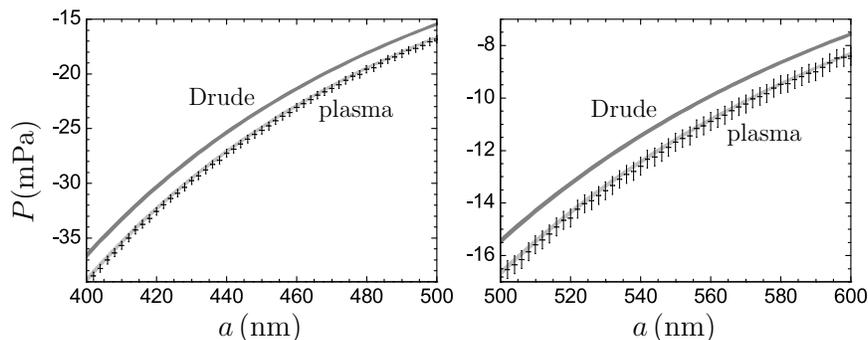}}
\vspace*{-11.5cm}
\caption{The effective Casimir pressure between two Au-coated plates computed
at $T=300~$K using either the plasma or Drude extrapolation of
the optical data is shown as a function of separation by the bottom
and top bands, respectively, over the intervals from 400 to 500~nm
(left) and from 500 to 600~nm (right). The measurement data are
shown as crosses.
\protect\label{figKM:3}}
\end{figure}

The roughness on the surfaces of both test bodies was measured by means of an atomic force
microscope and taken into account perturbatively. \cite{KM49,KM50}
In Fig.~\ref{figKM:3} we show two representative examples \cite{KM47} of the comparison
between experiment and theory where the bottom and top theoretical bands for the effective
pressure are computed at $T=300~$K by (\ref{eqKM3}), (\ref{eqKM5}), and the optical data
for Au extrapolated down to zero frequency using the plasma model (\ref{eqKM8}) and the
Drude model (\ref{eqKM7}), respectively. The experimental data are shown as crosses with
their total errors determined at the 95\% confidence level. It is seen that the theoretical
predictions using the plasma model extrapolation (\ref{eqKM8}) are in a good agreement with
the data whereas the Drude extrapolation (\ref{eqKM7}) is experimentally excluded.

According to Fig.~\ref{figKM:3}, the difference between two alternative theoretical
predictions using different extrapolations reaches only a few percent of the Casimir pressure
magnitude. This difference, however, can be increased significantly by measuring
the differential Casimir force. \cite{KM51,KM52}
Using this idea, measurements of the differential Casimir force between a Ni-coated sphere and
Au and Ni sectors of the disc coated with an Au overlayer have been performed by means of
a micromechanical torsional oscillator. \cite{KM53}
In the configuration of this experiment, the theoretical predictions using the plasma and
Drude extrapolations differ from one another by up to a factor of 1000. As a result, the
predictions of the Lifshitz theory using the plasma model extrapolation were found to be in
a good agreement with the measurement data whereas the alternative predictions using an
extrapolation of the optical data by means of the Drude model were conclusively excluded.

The next experiment was performed recently within a wider range of separations from 0.2 to
$8~\mu$m. In this case the micromechanical torsional oscillator was used to measure the
differential Casimir force between an Au-coated sphere and the top and bottom of Au-coated
deep Si trenches. \cite{KM54} An employment of the differential measurement scheme gave the
possibility to significantly decrease the total experimental errors (to below 3~fN at
separations exceeding $1~\mu$m). Due to the large deepness of the trenches, the force
between a sphere and their bottom vanishes. As a result, it is the Casimir force between
a sphere and a plane trench top which is measured. This work also differs from the discussed
above experiments in that it includes the characterization of patch potentials by means
of Kelvin probe microscopy and performs computations of the Casimir force in the sphere-plate
geometry based on the recently developed first-principle methods using the scattering
theory \cite{KM55,KM56,KM57,KM58} and the gradient expansion.  \cite{KM59,KM60,KM61,KM62}

\begin{figure}[b]
\vspace*{-8.5cm}
\centerline{
\includegraphics[width=6.50in]{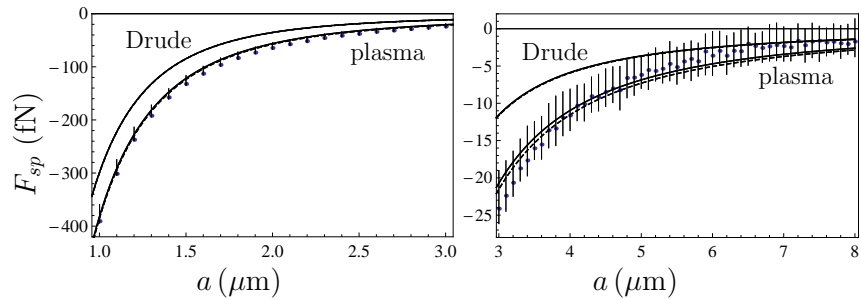}}
\vspace*{-11.2cm}
\caption{The Casimir force between an Au-coated sphere and an
Au-coated plate computed at $T=300~$K by the exact theory using
either the plasma or Drude extrapolation of the optical data is
shown as a function of separation by the bottom and top solid lines,
respectively, over the intervals from 1 to 3~$\mu$m (left) and from
3 to 8~$\mu$m (right). The respective PFA results are presented by
the dashed lines. The measurement data are shown as crosses.
\protect\label{figKM:4}}
\end{figure}
In Fig.~\ref{figKM:4} we present the comparison between experiment and theory over the
separation region from 1 to $8~\mu$m which is not covered in previous precision
experiments. \cite{KM54} The bottom and top solid lines are computed by the exact theory
using the plasma and Drude extrapolations of optical data, respectively. The theoretical
predictions obtained using the PFA are indicated by the dashed lines. They are not
discriminated from the solid lines in the limits of total experimental errors shown as crosses
which are determined at the 95\% confidence level. As can be seen in Fig.~\ref{figKM:4},
the theoretical predictions using the extrapolation by means of the Drude model are
experimentally excluded over the separation region $a<4.8~\mu$m. The predictions obtained
with the help of the plasma model extrapolation of the optical data are in agreement with
the data over the entire measurement range.

It is significant that the difference between two alternative theoretical predictions
in Figs.~\ref{figKM:3} and \ref{figKM:4} is entirely caused by the respective difference
in thermal corrections in Fig.~\ref{figKM:1} using the Drude and plasma extrapolations
of the optical data to zero frequency.

\subsection{Measurements by means of an atomic force microscope}
\label{subsecKM:3p2}

In four experiments, the gradient of the Casimir force between an Au-coated sphere and
an Au-coated plate $F_{sp}^{\,\prime}(a,T)$ was measured using a modified AFM operated
in the dynamic mode. \cite{KM63,KM64,KM65,KM66}
Similar measurements were also performed \cite{KM67,KM68,KM69} in configurations where either
a plate or both a sphere and a plate were coated with a magnetic (but not magnetized) metal Ni.
Using the PFA, the gradient of the Casimir force $F_{sp}^{\,\prime}$ was expressed via the
Casimir pressure for two parallel plates (\ref{eqKM2}) or
(\ref{eqKM5}) found in the framework of the Lifshitz theory
\begin{equation}
F_{sp}^{\,\prime}(a,T)=-2\pi R P(a,T).
\label{eqKM12}
\end{equation}

Small corrections to (\ref{eqKM12}) due to surface roughness and deviations from PFA were
taken into account perturbatively and contributed only a fraction of percent.
Much care was taken to diminish the role of electrostatic effects arising due to the residual
potentials and surface patches. Specifically, the setups of three
experiments \cite{KM64,KM65,KM66} were equipped with an Ar-ion guns and UV lamps allowing
to perform additional cleaning procedures of the sphere and plate surfaces.

\begin{figure}[b]
\vspace*{-8.5cm}
\centerline{
\includegraphics[width=6.50in]{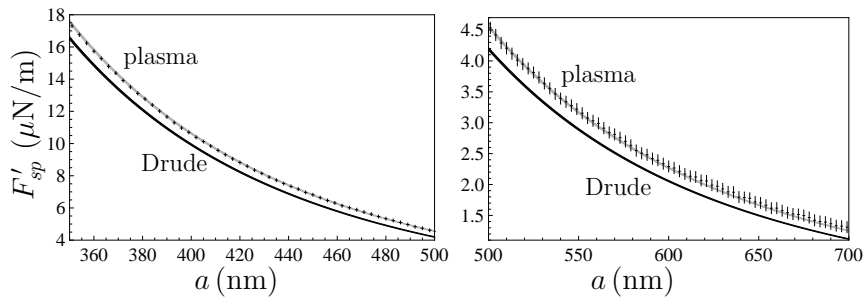}}
\vspace*{-11.2cm}
\caption{The gradient of the Casimir force between an Au-coated
sphere and an Au-coated plate computed at $T=300$K using
either the Drude or plasma extrapolation of the optical data is
shown as a function of separation by the bottom and top bands,
respectively, over the intervals from 350 to 500~nm (left) and from
500 to 700~nm (right). The measurement data are shown as crosses.
\protect\label{figKM:5}}
\end{figure}
In Fig.~\ref{figKM:5}, the typical results for the gradients of the Casimir force between
Au-coated test bodies measured in this way are shown as crosses where the total
experimental errors are determined at the 67\% confidence level. \cite{KM66}
The bottom and top theoretical bands are computed by (\ref{eqKM3}), (\ref{eqKM5}), and
(\ref{eqKM12}) where the optical data for Au were extrapolated using the Drude and plasma
models, respectively. It is again seen that the Lifshitz theory using the plasma model
extrapolation agrees with the measurement data whereas the same theory using the Drude model
extrapolation is excluded by the data.

An application of the atomic force microscope setup to the first measurements of the Casimir
interaction with magnetic test bodies \cite{KM67,KM68,KM69} was especially important for
the problem of thermal Casimir force. The reason is that for an Au-coated sphere above
a Ni-coated plate at the experimental separations of several hundred nanometers the Lifshitz
theory using the Drude and plasma extrapolations of the optical data leads to almost
coinciding theoretical predictions for the force gradients. \cite{KM67} As to the case of
both test bodies coated with the magnetic metal Ni, the force gradients computed using the
plasma model extrapolation turned out to be smaller than those computed using the Drude
model. \cite{KM68,KM69} This is on the contrary to the case of two nonmagnetic metals
(see Fig.~\ref{figKM:5}).

\begin{figure}[b]
\vspace*{-8.5cm}
\centerline{
\includegraphics[width=6.50in]{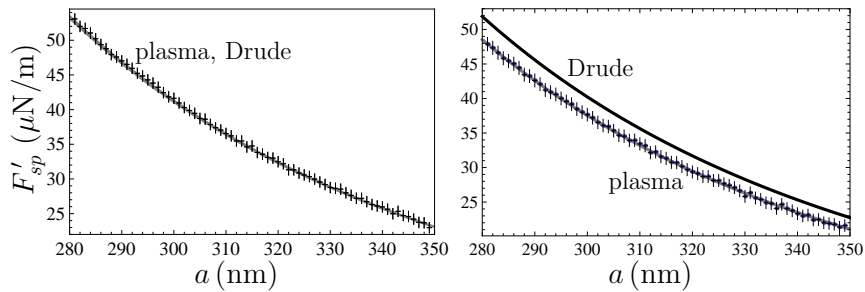}}
\vspace*{-11.2cm}
\caption{The gradient of the Casimir force between an Au-coated
sphere and a Ni-coated plate computed at $T=300$K using
either the Drude or plasma extrapolation of the optical data is
shown as a function of separation by the common band (left).
For the Ni-coated surfaces of a sphere and a plate, the force
gradient computed at $T=300$K using either the plasma or Drude
extrapolation is shown by the bottom and top bands,
respectively (right). In both cases the measurement data are
shown as crosses.
\protect\label{figKM:6}}
\end{figure}
Figure~\ref{figKM:6} (left) demonstrates the force gradients measured in the
experiment\cite{KM67} between an Au-coated sphere and a Ni-coated plate.
They are shown as crosses where the total experimental errors were determined
at the 67\% confidence level. These data are in good agreement with the theoretical
band which indicates the joint predictions obtained with either the Drude or the plasma
model extrapolation. In  Fig.~\ref{figKM:6} (right), the measured force gradients
between a Ni-coated sphere and a Ni-coated plate are presented as crosses. These data
are in a good agreement with the bottom theoretical band computed using the plasma
model extrapolation. The top band computed with the Drude model extrapolation is
excluded by the data.

The modified atomic force microscope was also used for testing the thermal Casimir force
with dielectric test bodies. Thus, the differential Casimir force $F_{\rm diff}$
between an Au-coated sphere and a Si membrane was measured in the presence and in the
absence of laser pulse on this membrane. \cite{KM70,KM71}
The membrane was made of $p$-type dielectric Si with the charge carrier concentration of
approximately $5\times 10^{14}~\mbox{cm}^{-3}$, i.e., much below the critical value at
which Si transforms into a metallic state. In the presence  of a laser pulse, the
concentration of  charge carriers jumped to $(1-2)\times 10^{19}~\mbox{cm}^{-3}$, i.e.,
an illuminated membrane was in the metallic state.

\begin{figure}[b]
\vspace*{-6.7cm}
\centerline{
\includegraphics[width=4.50in]{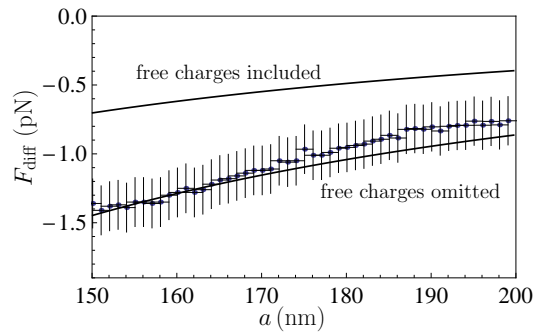}}
\vspace*{-5.3cm}
\caption{The differential Casimir force between an Au-coated
sphere and a Si membrane in the presence and in the absence of
laser pulse on its surface computed at $T=300~$K with either
omitted or included free charge carriers in a membrane in the
dark state is shown as a function of separation by the bottom
and top lines, respectively. The measurement data are shown as
crosses.
\protect\label{figKM:7}}
\end{figure}
In Fig.~\ref{figKM:7}, the measured difference of the Casimir forces in the presence and
in the absence of light is indicated
as crosses at different sphere-plate separations where the
experimental errors were determined at the 95\% confidence level. The bottom and top
theoretical lines were computed at $T=300~$K using the Lifshitz theory and the PFA with
the dielectric permittivity of membrane $\ve_{\rm opt}$ and $\ve$ from (\ref{eqKM9}),
respectively, in the dark (dielectric) state. The charge carriers in Au and in a Si
membrane in the presence of a laser pulse were described by the Drude model (in this
experiment, the use of the plasma model for Au and  for Si in the bright state leads to only
minor differences which cannot be discriminated in the limits of experimental errors).
As can be seen in Fig.~\ref{figKM:7}, the measurement data exclude the theoretical
predictions obtained with taken into account conductivity of a membrane in the dielectric
state. The same data are in agreement with theoretical predictions found with this
conductivity omitted.

Similar results have been obtained from measuring the Casimir force between an Au-coated
sphere and an indium tin oxide (ITO) film deposited on a quartz plate before and after
the UV irradiation of the film. \cite{KM72,KM73} It was observed that irradiation leads
to a significant decrease in the force magnitude measured with no detectable changes
in the optical data of the plate. This was interpreted as a phase transition of an
ITO film from a metallic to a dielectric state under the influence of UV irradiation.
In doing so, the measured Casimir force between an Au-coated sphere and an irradiated film agrees with the theoretical predictions only if the film material is described
by the permittivity
$\ve_{\rm opt}(\omega)$ with the role of free charge carriers omitted. \cite{KM72,KM73}

In the end of this section it is necessary to briefly discuss
an experiment on measuring the thermal Casimir-Polder
force between ${}^{87}$Rb atoms belonging to the Bose-Einstein condensate and a dielectric
fused silica plate. \cite{KM36} Although it is not performed by means of an AFM,
the obtained results are akin to those found for dielectric materials by means of an AFM.
In this experiment, the condensate cloud was resonantly driven into a dipole
oscillation by the magnetic field and the shift of the oscillation frequency caused by the
Casimir-Polder force was measured over the range of separations from 7 to $11~\mu$m.
The measurements were performed both in thermal equilibrium (when the plate temperature
$T_p=310~$K was equal to the environmental temperature) and out of thermal equilibrium
conditions ($T_p=479~$K and $T_p=645~$K). In the equilibrium conditions, the Casimir-Polder
force and respective frequency shift were computed by the Lifshitz theory using the
dielectric permittivity $\ve_{\rm opt}(\omega)$ of fused silica with neglected contribution
of nonzero conductivity. The computational results were found to be in a good agreement
with the measurement data. \cite{KM36}
If, however, computations of the Casimir-Polder force are performed with the permittivity
(\ref{eqKM9}) taking into account the conductivity of fused silica, the respective frequency
shifts are excluded by the data. \cite{KM74}

Thus, the measurement data of high precision
 Casimir experiments with dielectric test bodies agree with
theoretical predictions under a condition that the contribution of free charge carriers into
the dielectric permittivity is omitted.

\section{Theoretical Approaches to the Resolution of the Problem}
\label{secKM:4}

According to a large body of research presented above, theoretical
description of the thermal Casimir force faces with serious difficulties.
The Lifshitz theory of the Casimir force using the most realistic
response functions taking into account the relaxation properties of
conduction electrons in metals and free charge carriers in dielectrics
arrives at violation of the Nernst heat theorem for the plates
made both of an idealized metal with perfect crystal lattice and of a usual
dielectric material (for usual metallic plates the Nernst theorem is
satisfied due to the presence of some fraction of impurities).

What is more, the theoretical predictions of the Lifshitz theory using
the realistic response functions come into conflict with the measurement
data of all high precision experiments performed with both metallic and
dielectric test bodies. An agreement between experiment and theory is
restored when one neglects by the actually existing phenomena, such as
the relaxation of conduction electrons in metals and the presence of
free charge carriers in dielectrics at any nonzero temperature. This
situation cannot be considered as satisfactory. Below we briefly list
the main approaches proposed in the literature to remedy it.

\subsection{Validity of approximate methods}
\label{subsecKM:4p1}

Most of the experiments listed above (with exception of only
one \cite{KM54} performed in 2021) use the PFA in the comparison
between experiment and theory. Taking into account that until recently
the accuracy of PFA was unknown, it was suggested \cite{KM75} that
a disagreement between theory using the conventional response
functions and the measurement data may be explained by large errors
arising from an application of this approximation. Specifically,
using the generalization of the Lifshitz theory in the framework of
scattering approach, it was found that for an ideal metal sphere
above an ideal metal plate the exact results may depart from the PFA
ones. \cite{KM75}

Similar conclusions were made for usual metals \cite{KM76} but for
rather big values of the ratio $a/R>0.2$ (in typical experimental
conditions it holds $a/R<0.01$). By now the corrections to PFA in the
sphere-plate geometry are found with taken into account real material
properties by means of the Drude and plasma extrapolations of the
optical data using the scattering theory \cite{KM54,KM55,KM56,KM57,KM58}
and the gradient expansion. \cite{KM59,KM60,KM61,KM62} It was proven
that the PFA is well applicable for a comparison between experiment and
theory for the sufficiently small ratios of a separation to a sphere radius.

Another approximate method used in the theory-experiment comparison
was an additive approach to the surface roughness. \cite{KM49,KM50}
As was noted in the literature, the roughness correction should be
taken into account when comparing experiment with theory, especially
at the shortest separations between the test bodies. \cite{KM77,KM78}
Because of this, it is important to determine how accurate would
the calculated Casimir force be if the surface roughness is accounted
for additively.

This problem was solved within the scattering approach to the
Casimir force. According to the results obtained, \cite{KM79,KM80}
the additive approach is well applicable under a condition
$a \ll \Lambda_c$ where $\Lambda_c$ is the roughness correlation
length. Computations show that for a typical value of
$\Lambda_c=200~$nm the application region of the additive approach
is $a<2\Lambda_c/3 \approx 130~$nm. In this region, the roughness
correction may reach a few percent of the force magnitude even for
sufficiently smooth test bodies used in the Casimir experiments.
Under a condition $a>\Lambda_c$, the additive approach underestimates
the contribution of surface roughness to the Casimir
force. \cite{KM79,KM80} It turned out, however, that with increasing
$a$ the roughness correction decreases much faster than does the main
contribution to the force. Because of this, at separations where
the additive approach is inapplicable, one can simply neglect by the
roughness correction with no impact on the comparison between
experiment and theory.

\subsection{Analysis of the role of unaccounted effects}
\label{subsecKM:4p2}

It has been known that one difficulty which plagues measurements
of the Casimir interaction is an impact of the spatial distribution
of electrostatic potentials arising even on the grounded polycrystal
surfaces of a sphere and a plate in high vacuum. \cite{KM81} The
presence of these electrostatic effects, which are called the patch
potentials, results in the additional attractive force between a
sphere and a plate which increases the magnitude of the measured
force. Because of this, in the high precision experiments on
measuring the Casimir force special measures were undertaken in order
to keep the residual potential difference constant.

It was shown \cite{KM82} that the presence of sufficiently large
patches on the test body surfaces can compensate the differences
between top and bottom theoretical bands in Fig.~\ref{figKM:3} and,
thus, bring the theoretical predictions using the Drude model
extrapolation of the optical data in agreement with the measurement
results. However, direct measurements of the electrostatic potentials
on the test body surfaces by means of Kelvin probe microscopy
demonstrated \cite{KM83} that an additional force originating from
the surface patches is below 1\% of the predicted Casimir force and,
thus, is incapable to explain a difference between the top and
bottom bands in Fig.~\ref{figKM:3}.

Moreover, the experiments on measuring the Casimir interaction with
magnetic test bodies \cite{KM67,KM68,KM69} established that the
role of patch potentials in high precision experiments performed at
separations below 1$\mu$m is negligibly small. The point is that any
marked patch effect would introduce a disagreement between experiment
and theory in Fig.~\ref{figKM:6} (left) where the theoretical
predictions obtained using the Drude and plasma model extrapolations
are similar.

As to Fig.~\ref{figKM:6} (right) related to the case of two magnetic
test bodies, the presence of any patch effect would only increase a
disagreement between the measurement data and theoretical predictions
using the Drude model extrapolation. To bring the latter in
agreement with the data, one would need some additional repulsive
interaction which cannot be provided by the patch effect. Recall
also that the role of patch potentials can be diminished
significantly by performing the UV- and Ar-ion cleaning of the
surfaces. \cite{KM64,KM65,KM66} Thus, the problem of thermal Casimir
force is not caused by the patch potentials although their
contribution should be carefully taken into account in all high
precision experiments (especially in those performed at separations
exceeding 1~$\mu$m).

One more effect that makes an impact on the comparison between
experiment and theory is connected with a choice of the optical
data of interacting bodies used in computations. Several high
precision measurements employed the handbook data \cite{KM19}
of Au in order to calculate the Casimir force. It was noted,
however, that the optical properties of Au films vary for each
specific sample, depend on the method of deposition and are
characterized by different values of the plasma frequency and
relaxation parameter. \cite{KM84,KM85} This should influence
the Casimir force and, thus, the comparison between experiment
and theory.

An analysis of different sets of optical data for Au samples
available in the literature led to the conclusion, \cite{KM17,KM50}
however, that they bring to even larger departure of the theoretical
predictions using the Drude model extrapolation from the measurement
results than the handbook data. Furthermore, in several experiments
either the Drude parameters or the complete sets of optical data have
been measured by means of ellipsometers for the specific samples
used. \cite{KM47,KM48,KM54,KM72,KM73} In doing so, the especially
measured optical data did not deviate significantly from the
handbook ones with no impact on the comparison of experiment and
theory. This means that although it is always preferable to
measure the optical data of each specific sample in each successive
experiment, the use of handbook data \cite{KM19} cannot leave to
qualitatively incorrect conclusions when comparing experiment
with theory.

\subsection{The spatial dispersion and screening effects}
\label{subsecKM:4p3}
\newcommand{\Te}{{\varepsilon^{\rm Tr}}}
\newcommand{\Le}{{\varepsilon^{\,\rm L}}}

Strictly speaking, the Lifshitz theory is applicable to the plate
materials possessing the temporal dispersion. The respective
dielectric permittivities of plates depend only on the frequency.
The spatial dispersion is characterized by the two dielectric
permittivities, the transverse one $\Te(\omega, \mbox{\boldmath$k$})$
and the longitudinal one $\Le(\omega, \mbox{\boldmath$k$})$, which
describe the response of matter to the electric fields {\boldmath$E$}
perpendicular and parallel to the wave vector {\boldmath$k$},
respectively. \cite{KM20} Such permittivities can be introduced
only under a condition of space homogeneity which is violated
by the presence of two parallel plates separated with a gap.
However, in the approximation of specular reflection valid for
sufficiently smooth test bodies used in the Casimir experiments
it is possible to introduce the fictitious homogeneous medium
because an electron reflected on the vacuum gap-plate interface
is indistinguishable from an electron coming on the source side
of this medium. This makes possible \cite{KM87,KM88} to express
the reflection coefficients in the Lifshitz formula via the
surface impedances and finally via the permittivities
$\Te$ and $\Le$ (see Sec.~\ref{secKM:5}).

It has been known that for metals a connection between the electric
field and the current becomes nonlocal in the region of the
anomalous skin effect which extends over the range of frequencies
from approximately $10^{12}$ to $10^{13}~$rad/s at $T=300~$K. With
decreasing temperature, the left boundary of this region goes down.
The spatial dispersion also appears in the screening of static
electric field in materials which contain some fraction of
free charge carriers. In doing so, the electric field penetrates
into the conducting media to a depth of the screening length.
The specific expression for this depth depends on whether the
density of charge carriers in the plate material goes to zero
(dielectrics) or remains nonzero (metals) with vanishing
temperature (the Debye-H\"{u}ckel and Thomas-Fermi screening
lengths, respectively \cite{KM89}). It should be noted that both
the anomalous skin effect and the screening effects occur in real
electromagnetic fields (the propagating waves including the case
of zero frequency).

The spatially nonlocal dielectric permittivities of electron plasma
were calculated in the random phase approximation by means of the
Boltzmann transport theory. \cite{KM90,KM91} It was shown that
the most important difference from the local response function is
demonstrated by the longitudinal permittivity
$\Le(\omega, \mbox{\boldmath$k$})$. The obtained results and similar
in spirit approaches were employed to calculate the Casimir and
Casimir-Polder forces. \cite{KM87,KM92,KM93,KM94} The Lifshitz
theory using the reflection coefficient at zero frequency
accounting for the effects of screening was found \cite{KM92} in
agreement with measurements of the Casimir-Polder force at large
${}^{87}$Rb atom-silica plate separations. \cite{KM36} However, an application
of similar approach at all Matsubara frequencies \cite{KM93} was
as yet excluded \cite{KM95} by the results of experiment on
measuring the differential Casimir force between an Au-coated
sphere and a Si membrane illuminated with laser
pulses \cite{KM70,KM71} (see Sec.~\ref{subsecKM:3p2}).

In the case of metallic Casimir plates, an employment of nonlocal
dielectric functions describing the anomalous skin effect leads
to negligibly small corrections to the Casimir force computed
using the Drude model \cite{KM87} which vary from 0.3\% to 0.1\%
when the separation increases from 100 to 300 nm. Calculation of
the thermal Casimir pressure between metallic plates with account
of screening effects using the Thomas-Fermi length \cite{KM93,KM94}
leads to approximately the same results as the spatially local
Drude model. In both cases the obtained theoretical predictions
are excluded \cite{KM95} by the measurement data of many high
precision
experiments. \cite{KM45,KM46,KM47,KM48,KM53,KM54,KM63,KM64,KM65,KM66}

The consideration of spatial dispersion allowed to make some
progress regarding a violation of the Nernst heat theorem
(see Sec.~\ref{secKM:2}).
For metallic materials it was shown \cite{KM88,KM96}
that the spatial dispersion can play the same role as the
residual relaxation by making the Casimir entropy equal to zero
at zero temperature. It was also shown that with account of
screening effects the Nernst theorem is satisfied for
dielectrics whose conductivity vanishes with temperature
exponentially fast due to the vanishing concentration of
charge carriers. \cite{KM92,KM93} There are, however, dielectric
materials, such as dielectric-type semiconductors, dielectrics with
ionic conductivity and dielectric-type doped semiconductors, for
which the concentration of charge carriers does not vanish with
$T$ whereas the conductivity goes to zero due to the vanishing
mobility of charge carriers. For these dielectric materials
the problem of violation of the Nernst heat theorem for the
Casimir entropy remains unresolved. \cite{KM95}

Thus, although a consideration of the effects of spatial dispersion
in the anomalous skin effect and screening effects leads to some
encouraging results concerning a discrepancy of theoretical
predictions of the Lifshitz theory with thermodynamics and
experiment, it does not solve the problem of thermal Casimir force.

\section{Could an Alternative Nonlocal Response to Evanescent Waves Solve
the Problem?}
\label{secKM:5}
\newcommand{\VT}{{v^{\rm Tr}}}
\newcommand{\VL}{{v^{\,\rm L}}}

The spatially nonlocal dielectric functions discussed in Sec.~\ref{subsecKM:4p3}
describe the anomalous skin effect and screening effects observed in the propagating
electromagnetic waves of rather high frequency and quasistatic fields, respectively.
These functions were derived under the conditions $\omega\ll\omega_p$ and
$k\ll k_{\rm F}$ where $k_{\rm F}$ is the Fermi wave number. \cite{KM90}
As mentioned in Sec.~\ref{subsecKM:4p3}, the respective transverse response functions
are rather close to the local ones measured in the optical experiments performed in the
area of propagating waves whereas the longitudinal ones deviate significantly from
the local results.

It should be particularly emphasized that the longitudinal response function
$\Le(\omega,\mbox{\boldmath$k$})$ can be found not only from optical measurements but
from experiments on measuring the energy loss and momentum transfer of high-energy
electrons which belong to the beam passing through a thin metallic film. \cite{KM97}
As a result, from a fundamental standpoint, an experimental information about $\Le$
is available in both areas of the propagating and evanescent waves. There is, however,
no experimental evidence about $\Te$ in the area of evanescent waves. Because of this,
it might be unjustified to extrapolate the dielectric permittivities $\Te$ suitable
for theoretical description of the anomalous skin effect or screening effects to the
area of evanescent waves and substitute them to the Lifshitz formula (\ref{eqKM2})
in the frequency interval $0\leqslant\omega\leqslant c\kb$.

Below we consider the phenomenological nonlocal response functions which do not aim to
describe the anomalous skin effect or screening effects in real fields but provide some
alternative in the area of evanescent waves off the mass shell which could provide
a solution to the problem of thermal Casimir force. We begin with the representation
of reflection coefficients applicable in the presence of spatial dispersion.

\subsection{Reflection coefficients in the approximation of specular reflection}
\label{subsecKM:5p1}

It is well known that the reflection coefficients on the boundary plane $\rv_{\bot}=(x,y)$
of a material plate (the $z$ axis is perpendicular to it) can be expressed it terms of
the surface impedances defined as \cite{KM97}
\begin{equation}
Z_{\rm TM}\okb=\frac{E_x(+0;\omega,\kb)}{H_y(+0;\omega,\kb)},
\qquad
Z_{\rm TE}\okb=-\frac{E_y(+0;\omega,\kb)}{H_x(+0;\omega,\kb)},
\label{eqKM13}
\end{equation}
\noindent
where all fields have the form
\begin{equation}
\mbox{\boldmath$F$}(t,\rv)=\mbox{\boldmath$F$}(z;\omega,\kb)\,
e^{-i\omega t+i\mbox{\boldmath$k$}_{\bot}\mbox{\boldmath$r$}_{\bot}}.
\label{eqKM14}
\end{equation}
\noindent
Using (\ref{eqKM13}), the reflection coefficients are given by \cite{KM97}
\begin{equation}
r_{\rm TM}\okb=\frac{cq+i\omega Z_{\rm TM}\okb}{cq-i\omega Z_{\rm TM}\okb},
\qquad
r_{\rm TE}\okb=\frac{cq Z_{\rm TE}\okb+i\omega}{cq Z_{\rm TE}\okb-i\omega}.
\label{eqKM15}
\end{equation}

If the plate material possesses only the temporal dispersion, the impedances
(\ref{eqKM13}) are equal to \cite{KM98}
\begin{equation}
Z_{\rm TM}\okb=\frac{ick\okb}{\omega\ve(\omega)},
\qquad
Z_{\rm TE}\okb=-\frac{i\omega\mu(\omega)}{ck\okb},
\label{eqKM16}
\end{equation}
\noindent
where $k\okb$ is defined in (\ref{eqKM4}), and (\ref{eqKM15}) reduces to the standard
Fresnel reflection coefficients (\ref{eqKM3}).

If the plate material possesses the spatial dispersion, the expressions for surface
impedances can be derived in the approximation of specular reflection (see
Sec.~\ref{subsecKM:4p3}). They take the following form:
\begin{eqnarray}
Z_{\rm TM}\okb&=&\frac{i\omega c\mu(\omega)}{\pi}\int_{-\infty}^{\infty}
\frac{dk_3}{\vk^2}\left[\frac{\skb}{\Le(\omega,\vk)\mu(\omega)\omega^2}+
\frac{k_3^2}{\Te(\omega,\vk)\mu(\omega)\omega^2-c^2\vk^2}\right],
\nonumber \\
Z_{\rm TE}\okb&=&\frac{i\omega c\mu(\omega)}{\pi}\int_{-\infty}^{\infty}
\frac{dk_3}{\Te(\omega,\vk)\mu(\omega)\omega^2-c^2\vk^2},
\label{eqKM17}
\end{eqnarray}
\noindent
where $\vk=(k_1,k_2,k_3)$. For nonmagnetic materials ($\mu=1$) the expressions (\ref{eqKM17})
have long been derived by different methods. \cite{KM90,KM99}
The generalization for the case of magnetic materials was made very recently. \cite{KM100}

By putting $\omega=i\xi_l$ in (\ref{eqKM15}) and (\ref{eqKM17}), one can calculate the Casimir
pressure between the plates with account of spatial dispersion using the Lifshitz formula
(\ref{eqKM5}). This was made, for instance, for the nonmagnetic metal plates with account
of the anomalous skin effect \cite{KM87} but the results obtained were almost the same as
are given by the standard Drude model (see Sec.~\ref{subsecKM:4p3}).

\subsection{An alternative response to the evanescent waves}
\label{subsecKM:5p2}

Taking into account that an application of the standard spatially nonlocal response
functions does not solve the problem of the thermal Casimir force, it was
suggested \cite{KM65,KM66} that a model which describes well the response of metal to
electromagnetic field on the mass shell may fail in describing the response to quantum
fluctuations which are off the mass shell.

Following this line of attack, the spatially nonlocal Drude-like response functions were
introduced \cite{KM101}
\begin{eqnarray}
\Te\okb&=&1-\frac{\omega_p^2}{\omega[\omega+i\gamma(T)]}\left(1+
i\frac{\VT\kb}{\omega}\right),
\nonumber \\
\Le\okb&=&1-\frac{\omega_p^2}{\omega[\omega+i\gamma(T)]}\left(1+
i\frac{\VL\kb}{\omega}\right)^{-1},
\label{eqKM18}
\end{eqnarray}
\noindent
where $\VT$ and $\VL$ are the constants having a dimension of velocity which are of the
order of Fermi velocity $v_{\rm F}\sim 0.01c$.

The characteristic feature of the dielectric functions (\ref{eqKM18}) is that for the
electromagnetic waves on the mass shell in vacuum satisfying the condition
$\omega>c\kb$ (i.e., for the propagating waves) they nearly coincide with the Drude dielectric
permittivity (\ref{eqKM7}) because
\begin{equation}
\frac{v^{\rm Tr,L}\kb}{\omega}\sim \frac{v_{\rm F}}{c}\,\frac{c\kb}{\omega}<
\frac{v_{\rm F}}{c}\ll 1.
\label{eqKM19}
\end{equation}
\noindent
Thus, the functions (\ref{eqKM18}) take into account the relaxation properties of conduction
electrons as does the Drude model. However, in the frequency region
$0\leqslant\omega\leqslant c\kb$ (i.e., for the evanescent waves) these functions can
significantly depart from the Drude model leading to some new results in the theoretical
description of physical effects caused by the electromagnetic fluctuations off the mass shell
(an importance of the frequency region $\omega\ll kc$ for resolution of the problem of
thermal Casimir force was also discussed using the hydrodynamic approximation for the
response function \cite{KM102}).

Computations of the effective Casimir pressure and the gradient of the Casimir force were
made for Au-coated surfaces of a sphere and a plate in high precision experiments
performed by means of  a micromechanical torsional oscillator \cite{KM45,KM46,KM47,KM48}
and an atomic force microscope \cite{KM63,KM64,KM65,KM66} using the optical data of Au
and Eqs.~(\ref{eqKM5}), (\ref{eqKM12}), (\ref{eqKM15}), (\ref{eqKM17}), and
(\ref{eqKM18}). For $\VT=\VL=7v_{\rm F,Au}$, the results obtained \cite{KM101,KM103}
turned out to be in as good agreement with the measurement data as was reached earlier by
using the nondissipative plasma model (\ref{eqKM8}) (see Sec.~\ref{secKM:3}).

Similar computations were performed \cite{KM100} in the configuration of experiment
\cite{KM68,KM69} on measuring the gradient of the Casimir force between the sphere and
plate surfaces coated with a magnetic metal Ni. The computational results taking dissipation
into account by the dielectric functions (\ref{eqKM18}) were found in a good agreement
with the measurement data under the conditions $\VT=\VL=7v_{\rm F,Ni}$.

It is significant that both for Au and Ni test bodies only the first condition
$\VT=7v_{\rm F,Au(Ni)}$ is necessary for reaching good agreement with the measurement
data. As to the second parameter, $\VL$, it could vary in the interval from 0 to
$10v_{\rm F}$ with no impact on the measure of agreement between experiment and
theory. This means that, unlike the case of nonlocal effects occurring in the
propagating waves, in the off-the-mass-shell fields of quantum fluctuations the proper
description of a dielectric response may demand significant departure of the transverse
permittivity from its local form.

It was shown\cite{KM104} that the Casimir entropy calculated using the Drude-like
dielectric functions (\ref{eqKM18}) satisfies the Nernst heat theorem, i.e., goes to zero
with vanishing temperature for the perfect crystall lattices and also for lattices
with impurities. We recall that the Casimir entropy calculated using the standard Drude
model (\ref{eqKM7}) violates the Nernst heat theorem for metals with perfect crystal
lattices (see Sec.~\ref{subsecKM:2p1}). In the literature,\cite{KM105,KM106} this
violation was explained by an initiation of the correlated glassy state which is out
of thermal equilibrium. Furthermore, according to the results obtained,\cite{KM101}
the permittivities (\ref{eqKM18}) satisfy the Kramers-Kronig relations derived for
analytic functions with poles of the first and second order at zero frequency and do
not lead to contradictions with the measurement data of optical experiments exploiting
the propagating waves.

It is pertinent to stress that the Drude-like response functions (\ref{eqKM18}) depend
only on $\kb$. Taking into account, however, that the fictitious space discussed in
Sec.~\ref{subsecKM:4p3} is a 3D homogeneous manifold, it would be more general to
consider nonlocal functions depending on all the wave vector components. Based on this,
the simplest generalization of (\ref{eqKM18}) was suggested \cite{KM107}
\begin{eqnarray}
\Te(\omega,k)&=&1-\frac{\omega_p^2}{\omega[\omega+i\gamma(T)]}\left(1+
i\frac{\VT k}{\omega}\right),
\nonumber \\
\Le(\omega,k)&=&1-\frac{\omega_p^2}{\omega[\omega+i\gamma(T)]}\left(1+
i\frac{\VL k}{\omega}\right)^{-1},
\label{eqKM20}
\end{eqnarray}
\noindent
where $k=|\vk|=(\skb+k_3^2)^{1/2}$.

\begin{figure}[t]
\vspace*{-8.5cm}
\centerline{
\includegraphics[width=6.50in]{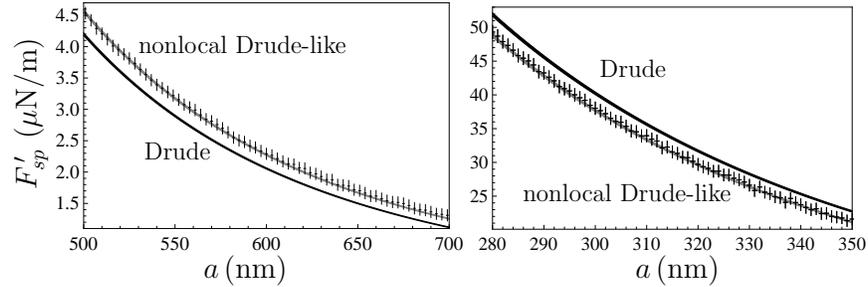}}
\vspace*{-11.2cm}
\caption{The gradient of the Casimir force between an Au-coated
sphere and an Au-coated plate computed at $T=300~$K using
either the Drude or nonlocal Drude-like extrapolations of the
optical data is shown as a function of separation by the bottom
and top bands, respectively (left). For the Ni-coated surfaces of
a sphere and a plate, the force gradient computed at $T=300~$K
using either the nonlocal Drude-like or Drude extrapolations is
shown by the bottom and top bands, respectively (right). In both
cases the measurement data are shown as crosses.
\protect\label{figKM:8}}
\end{figure}

Computations of the gradient of the Casimir force using the optical data
and Eqs.~(\ref{eqKM5}), (\ref{eqKM12}), (\ref{eqKM15}), (\ref{eqKM17}), and
(\ref{eqKM20}) were performed in the configurations of experiments exploiting the
Au-coated \cite{KM65,KM66} and Ni-coated\cite{KM68,KM69} surfaces of a sphere and
a plate at $T=300~$K with $\VT=\VL=3v_{\rm F,Au(Ni)}/2$.
The computational results for the Au-coated test bodies are shown in
Fig.~\ref{figKM:8} (left) by the top band and for the Ni-coated ones in
Fig.~\ref{figKM:8} (right) by the bottom band as the functions of separation.\cite{KM107}
The bottom band in Fig.~\ref{figKM:8} (left) and the top band in Fig.~\ref{figKM:8} (right)
are computed using the Drude model extrapolation of the optical data. The experimental
data are shown as crosses where the measurement errors were found at the 67\% confidence
level.

As is seen in Fig.~\ref{figKM:8}, the computational results using the Drude-like
response functions (\ref{eqKM20}) are in a good agreement with the measurement data.
The theoretical predictions using the Drude model are experimentally excluded.
By comparing Fig.~\ref{figKM:8} (left) with Fig.~\ref{figKM:5} (right) and
Fig.~\ref{figKM:8} (right) with Fig.~\ref{figKM:6} (right)  one can conclude that
the theoretical predictions of the Lifshitz theory using the nonlocal permittivities
(\ref{eqKM20}) taking into account the dissipation of free charge carriers are
indistinguishable from the predictions of the same theory made by using the
dissipationless plasma model. For the experiment employing dissimilar Au- and
Ni-coated test bodies \cite{KM67} an employment of  the nonlocal permittivities
(\ref{eqKM20})  with $\VT=\VL=3v_{\rm F,Au(Ni)}/2$ results in the theoretical band
coinciding with that in Fig.~\ref{figKM:6} (left) which is in agreement with the
measurement data.

Note that except of entirely theoretical advantage of the permittivities (\ref{eqKM20})
compared to (\ref{eqKM18}), they also lead to by a factor of 4.7 smaller values of the
constants $\VT$ and $\VL$ providing an agreement between experiment and theory in
measuring the Casimir force. This makes even smaller any corrections arising from
using (\ref{eqKM20}) in place of the standard Drude model (\ref{eqKM7}) in
interpretation of optical experiments performed in the area of propagating waves.
It is important also that the value of constant $\VL$ in (\ref{eqKM20}), as well as
in (\ref{eqKM18}), can vary in the interval from 0 to $10v_{\rm F,Au(Ni)}$ with
no impact on the measure of agreement between the theoretical Casimir forces and
the measurement data. This means that the major role in this agreement is
played by the Drude-like dielectric function $\Te(\omega,\vk)$. As to  $\Le(\omega,\vk)$,
it can be replaced even with the standard Drude model (\ref{eqKM7}).

\section{Conclusions and Outlook}
\label{secKM:6}

In the foregoing, we have considered the problem of thermal Casimir
force actively discussed in the literature for the last twenty years.
This problem lies in the fact that theoretical predictions of the
fundamental Lifshitz theory obtained using the universally accepted
dielectric functions are inconsistent with the measurement data of
high precision experiments and in some cases (idealized model of
metals with perfect crystal lattices, dielectrics with account of
electric conductivity at nonzero temperature) are found to be in
conflict with the Nernst heat theorem.

After presenting the results of numerous experiments on measuring
the thermal Casimir force performed by means of a micromechanical
torsional oscillator and an atomic force microscope, we have
outlined extensive studies of the problem directed to its resolution.
These include an investigation of the validity and application regions
of various approximate methods, including the PFA and the additive
approach to surface roughness, and an analysis of the role of possible
unaccounted effects, such as surface patches and sample-to-sample
variation of the optical data. Particular attention has been given
to a generalization of the Lifshitz theory with account of spatial
dispersion and screening effects. It is shown that although the
use of nonlocal response functions derived in the literature for
theoretical description of the anomalous skin effect and screening
effects leads to some encouraging results, the problem of
disagreement between experiment and theory remained unresolved.

Finally, the method of attack was considered which attracts
special attention to the fact that the transverse response
function to the evanescent waves cannot be determined
experimentally and its correct form may be unobtainable by the
analytic continuation of familiar nonlocal permittivities
describing, e.g., the anomalous skin effect. The recently
suggested phenomenological nonlocal Drude-like response functions
are discussed which take proper account of the dissipation of
conduction electrons and simultaneously bring the Lifshitz theory
in agreement with all high precision experiments on measuring the
Casimir force between metallic surfaces and with the Nernst heat
theorem.

In the future, it would be desirable to develop a complete
description of the dielectric response of both metallic and
dielectric materials based on first principles of quantum
electrodynamics at nonzero temperature. This could be attained
by deriving the polarization tensor of electronic plasmas in
dielectrics and metals like this is already done for graphene
in the framework of the Dirac model. \cite{KM108,KM109,KM110}
The first step in this direction is already made by finding
the polarization tensor of the three-dimensional Dirac
material. \cite{KM111} One may hope that the fundamental
derivation of response functions will make it possible not
only to obtain familiar results describing the anomalous skin
effect and screening effects as the special cases in
the area of propagating waves, but also justify the
phenomenological permittivities proposed in Sec. 5 as the
asymptotic results in the area of evanescent waves. In our
opinion, this is a plausible way toward final resolution of
the problem of thermal Casimir force.

\section*{Acknowledgments}

This work was partially
supported by the Peter the Great Saint Petersburg Polytechnic
University in the framework of the Russian state assignment for basic research
(project No.\ FSEG-2020-0024).
This paper has been supported by the  Kazan Federal University
Strategic Academic Leadership Program.

\end{document}